\documentclass[amsmath,amssymb,prl,hyperlink,twocolumn]{revtex4-2}

\usepackage{graphicx}
\usepackage{soul}
\usepackage[colorlinks=true,citecolor=red,linkcolor=magenta]{hyperref}
\usepackage[usenames]{color}
\usepackage{amsfonts}
\usepackage{color}
\usepackage{booktabs}
\usepackage{multirow}
\usepackage{float}
\usepackage{times}
\usepackage[english]{babel}
\usepackage{float}
\usepackage{caption}
\usepackage{ragged2e}

\begin{document}

\captionsetup[figure]{labelfont={bf},labelformat={default},labelsep=period,name={Fig}}
	
\title{Realization of fractional quantum Hall state with interacting photons}

\author{Can Wang$^{1,2}$}
\thanks{C. W., F.M. L., and M.C. C. contributed equally to this work.}
\author{Feng-Ming Liu$^{1,2}$}
\thanks{C. W., F.M. L., and M.C. C. contributed equally to this work.}
\author{Ming-Cheng Chen$^{1,2}$}
\thanks{C. W., F.M. L., and M.C. C. contributed equally to this work.}
\author{He Chen$^{1,2}$}
\author{Xian-He Zhao$^{1,2}$}
\author{Chong Ying$^{1,2}$}
\author{Zhong-Xia Shang$^{1,2}$}
\author{Jian-Wen Wang$^{1,2}$}
\author{Yong-Heng Huo$^{1,2,3}$}
\author{Cheng-Zhi Peng$^{1,2,3}$}
\author{Xiaobo Zhu$^{1,2,3}$}
\author{Chao-Yang Lu$^{1,2,3}$}
\email{cylu@ustc.edu.cn}
\author{Jian-Wei Pan$^{1,2,3}$ \vspace{0.4cm}}
\email{pan@ustc.edu.cn}
	
\affiliation{$^1$ Hefei National Research Center for Physical Sciences at the Microscale and School of Physical Sciences, University of Science and Technology of China, Hefei 230026, China}
\affiliation{$^2$ Shanghai Research Center for Quantum Science and CAS Center for Excellence in Quantum Information and Quantum Physics, University of Science and Technology of China, Shanghai 201315, China}
\affiliation{$^3$ Hefei National Laboratory, University of Science and Technology of China, Hefei 230088, China}
	
\begin{abstract}
Fractional quantum Hall (FQH) states, known for their robust topological order and the emergence of non-Abelian anyons, have captured significant interest due to the appealing applications in fault-tolerant quantum computing. Bottom-up approach on an engineered quantum platform will provide opportunities to operate FQH states without external magnetic field and enhance local and coherent manipulation of these exotic states. Here we demonstrate a lattice version of photon FQH state using a programmable on-chip platform based on photon blockade and engineering gauge fields on a novel two-dimensional circuit quantum electrodynamics (QED) system. We first observe the effective photon Lorentz force and butterfly spectrum in the artificial gauge field, a prerequisite for FQH states. After adiabatic assembly of Laughlin FQH wavefunction of 1/2 filling factor from localized photons, we observe strong density correlation and chiral topological flow among the FQH photons. We then verify the unique features of FQH states in response to external fields, including the incompressibility of generating quasiparticles and the smoking-gun signature of fractional quantum Hall conductivity. Our work represents a significant advance in the bottom-up creation and manipulation of novel strongly correlated topological quantum matter  composed of photons and opens up possibilities for fault-tolerant quantum information devices.  
\end{abstract}
	
\maketitle

	When charged particles are confined in a two-dimensional (2D) layer and subjected to an intense magnetic field at low temperatures, a new state of matter with remarkable properties, FQH states \cite{1982tsuiTwoDimensionalMagnetotransportExtreme}, will emerge. These states marry topological and strong correlated properties, and host quasiparticles with fractional charge and fractional statistics, showing far-reaching implications in the fields of condensed matter physics \cite{,2017wenColloquiumZooQuantumtopological,2019wenChoreographedEntanglementDances} and quantum information science \cite{,2003kitaevFaulttolerantQuantumComputation,2008nayakNonAbelianAnyonsTopological}. Beyond the realm of continuous 2D systems \cite{2020clarkObservationLaughlinStatesa}, in discrete lattice systems with strong particle interactions and effective magnetic flux, analogous fractionalization phenomena are also expected to occur as fractional Chern insulators \cite{2023leonardRealizationFractionalQuantum} without external magnetic field \cite{,2023caiSignaturesFractionalQuantum,2023xuObservationIntegerFractional}, which has sparked numerous theoretical studies on the creation and detection of FQH states in artificial lattice systems \cite{2018raciunasCreatingProbingManipulatingb, 2020repellinFractionalChernInsulatorsa,2022wangMeasurableSignaturesBosonicc,2005sorensenFractionalQuantumHalla,2007hafeziFractionalQuantumHalla,2008choFractionalQuantumHall,2012umucalilarFractionalQuantumHallb,2018raciunasCreatingProbingManipulatingb,2020repellinFractionalChernInsulatorsa,2022wangMeasurableSignaturesBosonicc,2022weberExperimentallyAccessibleScheme,2023wangColdatomElevatorEdgestate,2023liuRecentDevelopmentsFractional,2014kapitInducedSelfStabilizationFractionala,2023palmGrowingExtendedLaughlin}.
	
	Engineered quantum platforms offer a bottom-up approach to the experimental realization of such artificial lattice systems, including using ultracold atoms \cite{,2023leonardRealizationFractionalQuantum,2014jotzuExperimentalRealizationTopological,2015aidelsburgerMeasuringChernNumber,2015manciniObservationChiralEdgea,2017taiMicroscopyInteractingHarpera,2019cooperTopologicalBandsUltracold} or circuit QED techniques \cite{,2017guMicrowavePhotonicsSuperconducting,2020carusottoPhotonicMaterialsCircuit,2021blaisCircuitQuantumElectrodynamics}. Among these platforms, photons in circuit QED lattices stand out as a highly promising candidate due to their high on-chip scalability, great flexibility in lattice geometry \cite{2019kollarHyperbolicLatticesCircuit}, and the ability of universal control over individual sites. Numerous pioneering experimental efforts are devoted to the circuit QED platform \cite{2017roushanChiralGroundstateCurrents,2018owensQuarterfluxHofstadterLattice,2022owensChiralCavityQuantum,2023xiangSimulatingChernInsulators}, yet a simultaneous realization of photon interactions and artificial gauge fields \cite{2010kochTimereversalsymmetryBreakingCircuitQEDbased,2011nunnenkampSyntheticGaugeFields} on a sufficiently large 2D lattice remains a major challenge. Here we design and realize a novel circuit QED system using Plasmonium lattice to create and manipulate photon FQH states, and demonstrate the smoking-gun signature of fractional quantum Hall conductivity.

	\begin{figure*}[htb]
		\centering
		\includegraphics[width=0.81\textwidth]{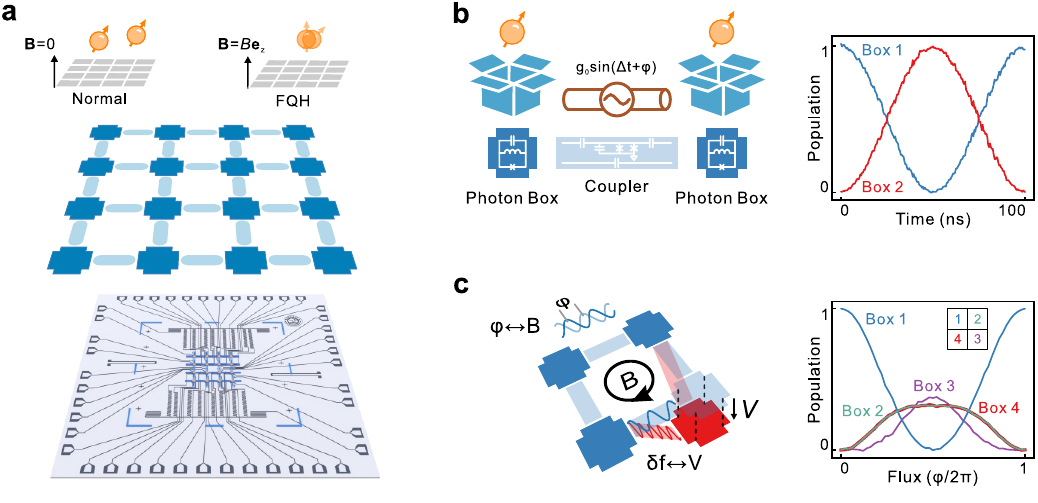}
		\caption{\justifying{\textbf{Circuit QED lattice for photon FQH states.} \textbf{a,} The circuit QED lattice used to implement photon blockade and artificial magnetic field, which consists of 16 Plasmonium photon boxes and 24 Floquet-driven couplers. \textbf{b,} A photon hops periodically between two adjacent boxes. When the coupler is driven on resonance with the frequency difference of the two boxes, the oscillation amplitude is maximized. \textbf{c,} Aharonov–Bohm effect. A photon hops in a four-site loop for a period of $\pi / 2|J|$. An interference pattern occurred depending on the effective magnetic flux $\phi$, which is proportional to the sum of the driving phases of the couplers along the loop as $\phi = \varphi/2\pi$. Disorder in the potential of a site can be introduced by detuning the driving frequencies of the couplers adjacent to the site (red site).}}
		\label{Fig1}
	\end{figure*}
	
	\begin{figure*}[htb]
		\centering
		\includegraphics[width=0.81\textwidth]{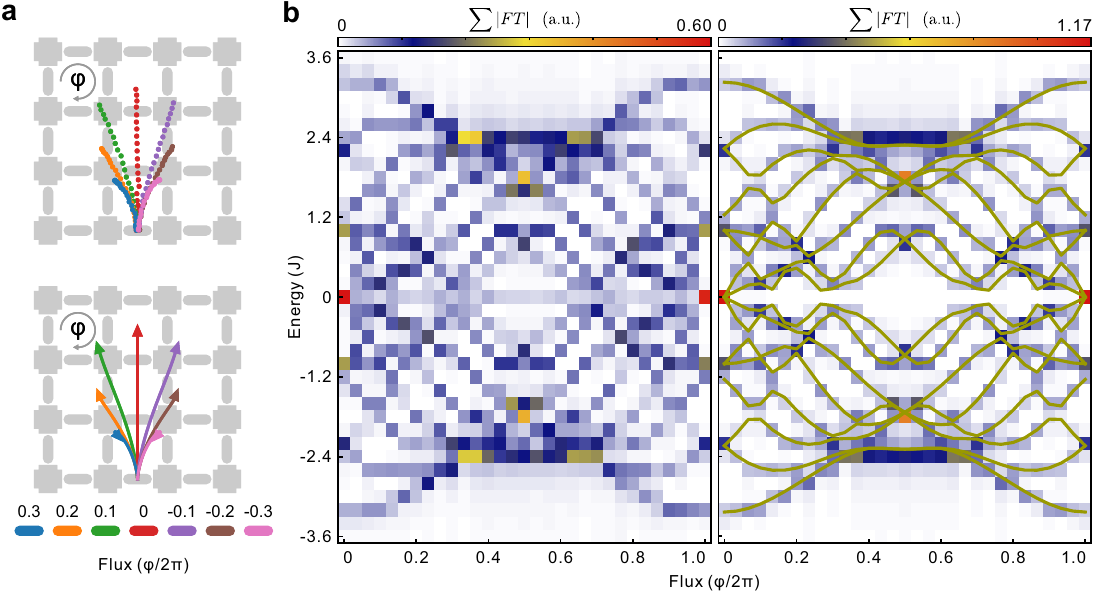}
		\caption{\justifying{\textbf{Verification of the synthesis of an artificial magnetic field.} \textbf{a,} The motion of a single photon is deflected by the effective Lorentz force induced by the artificial magnetic field. Top panel: experiment; Bottom panel: simulation. \textbf{b,} Hofstadter’s butterfly spectrum of a single photon is observed with respect to the magnetic flux through a single plaquette. Left panel: experiment; Right panel: simulation. The yellow solid lines represent the exact values obtained from diagonalization.}}
		\label{Fig2}
	\end{figure*}

	\textbf{Circuit QED lattice.} We use the recently developed Plasmonium photon box to design the FQH system \cite{2023liuQuantumComputeraidedDesign}, which is a Josephson-junction shunted superconducting oscillators with substantial anharmonicity. We extend it to form a two-dimensional lattice on a chip through a flip-chip assembly method (Fig.~\ref{Fig1}a). When a first photon hops into an empty box through plasmon transition, a second photon will be blocked due to the anharmonic energy penalty. The photon hopping is implemented through a Floquet driven coupler between two adjacent boxes \cite{,2016mckayUniversalGateFixedfrequency,2017eckardtAtomicQuantumGases}. We arrange 16 photon boxes and 24 couplers on a $4  \times  4$ lattice. The hopping amplitudes and phases can be dynamically programmed. Fig.~\ref{Fig1}b demonstrates the periodic hopping of a photon between two isolated sites when the Floquet driven is on resonance. Fig.~\ref{Fig1}c presents a clear phase interference pattern within a closed loop, which reproduces the Aharonov–Bohm effect \cite{2012fangPhotonicAharonovBohmEffect}.
	
	We lock the 24 hopping phases to each other and program them according to the Landau gauge to create an effective magnetic field on the 2D lattice \cite{1955harperGeneralMotionConduction,1976hofstadterEnergyLevelsWave}. Considering the significant photon blockade effect deriving from the anharmonicity of $400 \sim 470 \text{ }\rm{MHz}$, which is much greater than the maximum hopping rate $J / 2 \pi = 5 \text{ }\rm{MHz}$, we can express the effective system Hamiltonian in the two-dimensional subspace of each site $(x,y)$ as follows:
	\begin{equation}
		\hat{H}/\hbar = J\sum_{x,y}{\hat{\sigma}_{x,y+1}^\dag \hat{\sigma}_{x,y} + \hat{\sigma}_{x+1,y}^\dag \hat{\sigma}_{x,y} e^{-i2\pi\phi y} + \text{H.c.}}
	\end{equation}
	Here, $\hat{\sigma}^\dag$ ($\hat{\sigma}$) represents the creation (annihilation) operator of photons, and $\phi$ denotes the number of magnetic flux quanta in a single plaquette, which is proportional to the sum of hopping phases around the plaquette as $\phi = \varphi/2\pi$. The photon FQH state appears as the ground state of the Hamiltonian at an appropriate filling factor, which refers to $v=N/N_{\phi}$ with $N$ is the total number of photons and $N_{\phi}$ is the total number of magnetic flux quanta that thread through the whole system. Considering the exchange symmetry of many-body wavefunction, the bosonic FQH states can be well described by Laughlin wavefunctions at filling factors of even denominator, such as 1/2, in contrast to fermionic FQH states of odd denominator \cite{1983laughlinAnomalousQuantumHall}.
	
	\begin{figure*}[htb]
		\centering
		\includegraphics[width=0.85\textwidth]{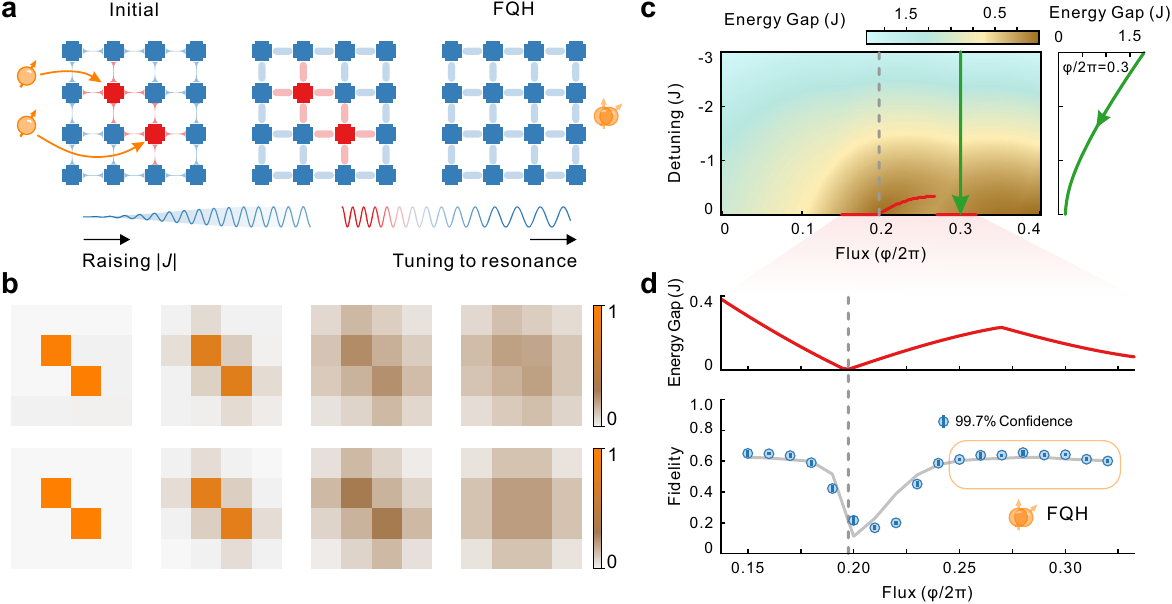}
		\caption{\justifying{\textbf{Adiabatic preparation of photon FQH states.} \textbf{a,} A disorder-assisted adiabatic protocol is used to prepare the system ground states. The photons are initially localized and isolated in the center of the lattice with large potential disorders (red sites). Subsequently, the couplings are ramped up and then potential disorders are flattened gradually. \textbf{b,} The photon density distribution across the lattice is monitored during the adiabatic evolution. Top panel: experiment; Bottom panel: simulation. \textbf{c,} Energy gaps are shown as a function of magnetic flux and potential disorders. The red line indicates the position of the minimum gap at varying magnetic fluxes. The green line represents the typical adiabatic path of our protocol. \textbf{d,} Topological phase transition. Ground state fidelities are measured with respect to the magnetic flux $\phi$. The fidelities are breakdown at the phase transition point, where the energy gap closes at $\phi=0.2$. The range of magnetic flux corresponding to the FQH regime is marked. The gray solid line represents the simulation result with decoherence.}}
		\label{Fig3}
	\end{figure*}
	
	\textbf{Photon Lorentz force and butterfly spectrum.} We first confirm the successful synthesis of an artificial magnetic field in the 2D lattice by monitoring the motion of a single photon, which could be deflected by the effective Lorentz force induced by the artificial magnetic field. We initialize a photon in a two-site Bell superposition state at the bottom of the lattice, as shown in Fig.~\ref{Fig2}a. In the absence of magnetic field, the centroid of the photon follows a straightforward upward trajectory during its evolution. However, when the magnetic field increases, such spatial inversion symmetry is broken and the photon is deflected to the right (left) at negative (positive) magnetic field (Fig.~\ref{Fig2}a).
	
	Additionally, we verify the Hofstadter butterfly energy spectrum, a phenomenon arising from the interplay between the artificial magnetic field and the lattice structure \cite{1955harperGeneralMotionConduction,1976hofstadterEnergyLevelsWave}. As the strength of the magnetic field varied, the energy spectrum of the system exhibits a fractal structure. We initialize one of lattice sites in a Fock superposition state consisting of vacuum and single photon, and then track its dynamic evolutions \cite{,2015jurcevicSpectroscopyInteractingQuasiparticles,2017roushanSpectroscopicSignaturesLocalization} (see Methods). We observe the Fourier spectrum displayed a distinctive butterfly-like pattern, which matched the exact values obtained through diagonalization (Fig.~\ref{Fig2}b).

	\textbf{Adiabatic preparation of photon FQH states.} We employ a disorder-assisted adiabatic protocol to prepare the photon FQH states. The protocol begins by initializing the system in a simple ground state, where $N  =  2$ photons (see Fig.~\ref{exfig1} for $N  =  3$) were localized at the center of the lattice. The effective potentials of these $N$ sites are significantly detuned down ($2\pi  \times  15 \text{ } \rm{MHz}$), and their couplings to the rest of the system are turned off (Fig.~\ref{Fig3}a and \ref{Fig3}b). Subsequently, the protocol gradually evolves the system by ramping up the couplings $J$ (to $2\pi  \times  5 \text{ } \rm{MHz}$) and flattening the potential disorders to reach the final Hamiltonian $H$ (Fig.~\ref{Fig3}c).
	
	During a slow evolution of $1 \text{ } \mu \rm{s}$, the system follows the instantaneous ground state, except for the Hamiltonian $H$ of magnetic flux near $\phi=0.2$, where a topological phase transition occurs. The ground state is topologically trivial for $\phi < 0.2$, while it becomes topologically nontrivial for $\phi > 0.2$. In our finite-size lattice, the Laughlin wavefunction with a filling factor 1/2 can be stabilized within a range of $\phi$ from 0.25 up to 0.32, as indicated in Fig.~\ref{Fig3}d. 
	
	To evaluate the fidelity $F$ of the prepared states, we subject them to the exact reversed adiabatic path and measure the probability of the photons returning to the initial localized state, represented by $F^2$ (Fig.~\ref{Fig3}d). The extracted fidelity of the FQH state, which is approximately 60\%, is in accordance with the level of decoherence in numerical simulations (96\% without decoherence). Additionally, we observe a fidelity breakdown at $\phi = 0.2$, which coincided with the gapless phase transition point in the theoretically calculated energy spectrum.
	
	\begin{figure*}[htb]
		\centering
		\includegraphics[width=0.85\textwidth]{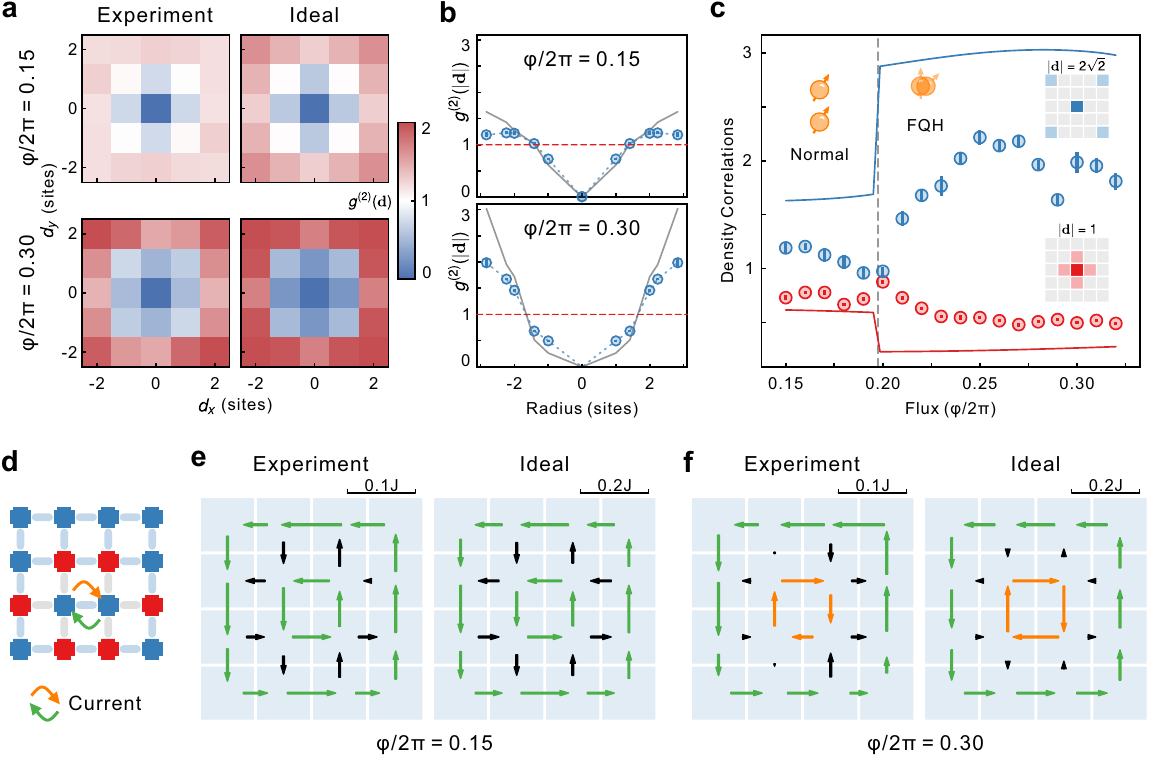}
		\caption{\justifying{\textbf{Characterization of the ground-state wavefunctions.} \textbf{a,} The density correlation functions $g^{(2)}(\bf{d})$ with respect to the relative position of photons. Top panel: the normal state; Bottom panel: the FQH state. \textbf{b,} The averaged density correlations as a function of photon radial distance. The gray solid lines represent the exact values obtained from diagonalization. \textbf{c,} The density correlations of small and large distances at varying magnetic flux. In the transition from the normal states to the FQH states, density correlations are suppressed at small distances while being enhanced at larger distances. The solid lines represent the exact values obtained from diagonalization. \textbf{d,} Measurement of the density current through the velocity of photon exchange across two isolated sites. \textbf{e,} The chiral density currents of the normal state. Strong radial currents flow between the boundary and the bulk. \textbf{f,} The chiral density currents of the FQH state. The radial currents are significantly suppressed.}}
		\label{Fig4}
	\end{figure*}
	\begin{figure*}[htb]
		\centering
		\includegraphics[width=0.85\textwidth]{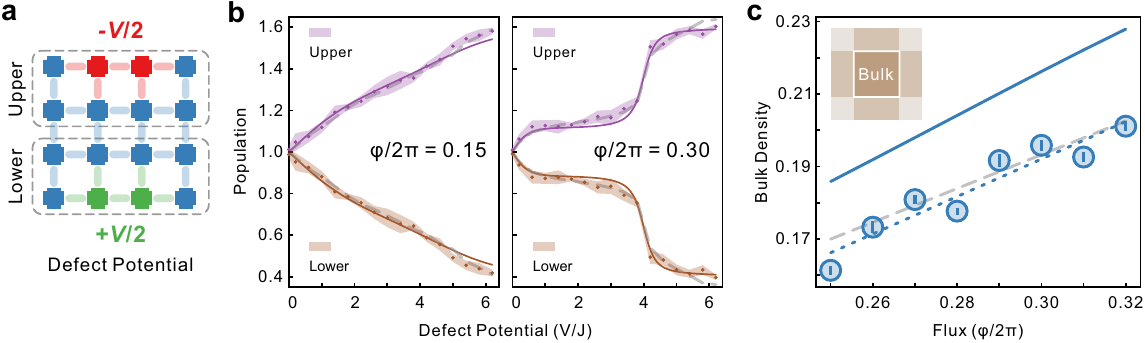}
		\caption{\justifying{\textbf{Response to external fields.} \textbf{a,} Local potential defects are introduced to excite quasiparticles and quasihole. Potential dips and bumps are highlighted in red and green, respectively. The changes in the photon number in the upper and lower sections are monitored. \textbf{b,} The measured changes in photon number as a function of the depth of the potential defects (99.7\% confidence interval). Left: the normal state; Right: the FQH state. In contrast to the linear response of the normal state, the FQH state exhibits a significant suppression of changes in photon number at shallow defects ($V/J<3$). The solid lines represent the simulation results without decoherence, scaled by a factor of 0.6 to guide the eye. The gray dash lines represent the simulation results with decoherence. \textbf{c,} The response of the bulk density of the photon FQH state to the magnetic flux. The data shows a linear trend fitted by the blue dashed line with a slop of 0.52(14), indicating a quantum Hall conductivity of $0.52(14)\sigma_{0}$. The blue solid (gray dashed) line represents the simulation result without (with) decoherence.}}
		\label{Fig5}
	\end{figure*}
	
	\textbf{Characterization of the ground-state wavefunctions.} In a FQH system, while particles are confined within the lowest degenerate levels with fixed kinetic energies, they minimize their repulsive potential by orbiting around each other to reach the ground state. This behavior imprints unique spatial characteristics to the wavefunction, which allow us to observe pattern differences, such as density correlation \cite{,2018raciunasCreatingProbingManipulatingb} and density currents \cite{2023tranMeasuringArbitraryPhysical}, between the normal state (at $\phi=0.15$) and the photon FQH state (at $\phi=0.3$). 
	
	In Fig.~\ref{Fig4}a, we examine the density-density correlation function $g^{(2)}(\bf{d})$, with respect to the relative position of photons $\bf{d}$ (see Methods). We observe that at small distances, photons exhibit antibunching behavior ($g^{(2)}<1$) due to photon blockade. Conversely, at large distances, the FQH state displays strong correlations of photons ($g^{(2)}>1$), which are significantly stronger than those observed in the normal state (Fig.~\ref{Fig4}b). We further measure the density correlations at different magnetic flux across the phase transition point $\phi=0.2$ (Fig.~\ref{Fig4}c). The results clearly reveal that during the transition from normal states to FQH states, photon correlations are suppressed at small distances while being enhanced at larger distances. This observation evidences a long-range repulsive pattern of photon pairing in FQH states.
	
	As the photons are expected to propagate in pairwise rotations, we go to investigate the topological pattern of photon motion by measuring the photon density current. We derive these values from the velocity of photon exchange across two isolated lattice sites (Fig.~\ref{Fig4}d and also see Methods). In the normal state (Fig.~\ref{Fig4}e), we observe anticlockwise chiral currents flowing along the boundary and also in the bulk, accompanied by strong radial currents flowing between the boundary and the bulk. However, in the FQH state (Fig.~\ref{Fig4}f), reversed currents emerge in the bulk, while the radial currents are significantly suppressed. This separation of the bulk and the boundary offers an advantage for further investigating the response of the FQH state in the bulk to external fields, where the boundary photons can serve as a particle reservoir for the bulk.
	
	\textbf{Observation of fractional response to external fields.} After using the correlation functions and density currents as initial evidence to characterize the prepared states, we turn to probe the hallmark response of FQH states to external fields. We first investigate the incompressibiltiy of generating quasiparticles above the many-body energy gap on the ground state. In order to investigate this behavior, we control the potential landscape of the FQH system by introducing local potential defects on the lattice to excite and trap the quasiparticles \cite{,2018raciunasCreatingProbingManipulatingb,2022wangMeasurableSignaturesBosonicc}. Fig.~\ref{Fig5}a shows the potential dips and bumps of depth $V/2$ are localized at the upper and lower boundaries of the lattice, which act as quasiparticle and quasihole traps, respectively. The generation process is monitored by examining the imbalance of the total photon number on the upper and lower sections of the lattice (Fig.~\ref{Fig5}b).

	In normal states (at $\phi=0.15$), the change in photon number accumulates linearly with the depth of local potential $V$. However, in FQH states (at $\phi=0.3$), the change in photon number is suppressed at shallow defects ($V/J<4$). This suppression clearly indicates the incompressibility of the FQH liquid due to the energy gap (Fig.~\ref{Fig5}b). As the potential continues to increase, the photons rapidly transfer from the lower to the upper section, corresponding to the formations quasiparticle and quasihole. As a result, we can observe an quick increase in photon change near $V/J=4$, followed by a subsequent suppression until photon saturation is reached.
	
	Another smoking-gun signature of FQH states is the existence of fractional quantum Hall conductivity $\sigma_{H}$. Interestingly, for the present incompressible states, the quantum Hall conductivity can be described using Středa’s formula \cite{,1982stredaTheoryQuantisedHall,2020repellinFractionalChernInsulatorsa}. According to this formula, $\sigma_{H}$ is proportional to the variation of bulk density of the FQH state with respect to the magnetic flux $\phi$. Here, the bulk density refers to the average photon number of the central four sites of the lattice.
	
	In Fig.~\ref{Fig5}c, we observe that the bulk density exhibits an approximately linear response to changes in the effective magnetic flux within the FQH regime, with a slope of 0.52(14). It indicates that the value of the quantum Hall conductivity is $0.52(14)\sigma_{0}$, where $\sigma_{0}$ is the inverse of the von Klitzing resistance $h/{2e}$. This result is consistent with the ideal value of 1/2 for half-filling FQH states, as well as with the value of 0.6 in numerical simulation that consider decoherence and finite-size effect.
	
	\textbf{Outlook.} In summary, we have realized a lattice version of photon FQH states, a long-sought goal in topological photonics over the past decade \cite{2019ozawaTopologicalPhotonics}, on a novel 2D circuit QED lattice with single-site programmability, and obtained definitive signatures verifying the exotic correlated topological properties. The precise control of quantum nonlinear optics at single photon level \cite{,2014changQuantumNonlinearOpticsb,2018changQuantumMatterBuilt} together with the inherent scalability of integrated QED circuits provides the present approach great potential for exploring new strongly interacting topological quantum matter made of photons \cite{2013carusottoQuantumFluidsLighta,2019maDissipativelyStabilizedMotta,2022saxbergDisorderassistedAssemblyStrongly,2022dengObservingQuantumTopology} and practical applications in topological quantum computing.
	By expanding the system size, the assembly of large-scale FQH states \cite{,2014kapitInducedSelfStabilizationFractionala,2023wangColdatomElevatorEdgestate,2023palmGrowingExtendedLaughlin} will further unlock the advantage of single-site programmability for topological coherent movement of localized anyons; and by  engineering tailored multi-photon interactions beyond photon-photon blockade \cite{,2014hafeziEngineeringThreebodyInteraction,2013kapit4bodyInteractions2body,2022zhangSynthesizingFiveBodyInteraction}, it will open up a realm to synthesize and control more exotic emergent particles with non-Abelian statistics as fault-tolerant quantum information hardware \cite{2008nayakNonAbelianAnyonsTopological}.
	
\begin{acknowledgments}

We thank Xiaodong Xu and Yao Wang for helpful discussions.

\end{acknowledgments}
	
	~\\
	
	\bibliographystyle{apsrev4-2}
	\bibliography{FQH.bib}
	
	~\\
	~\\
	\textbf{Methods} 
	~\\
	
	\textbf{Device information.} The circuit QED lattice is assembled using flip-chip technology. The top chip, referred to as the quantum chip, contains a Plasmonium array and Transmon-based tunable couplers that connect them. The bottom chip includes classical signal interfaces for control and readout. The Plasmonium is a compact photon box that composed of two metallic Al islands (charging energy $E_C/h \approx 0.60 \text{ }\rm{GHz}$) connected with a superconducting inductance (inductive energy $E_L/h \approx2.70 \text{ }\rm{GHz}$ ), and shunted by a nonlinear Josephson junction (Josephson energy $E_J/h\approx 4.58 \text{ }\rm{GHz}$). The photon anharmonicity exceeds 400 MHz, and the average photon decay time is 28 $\mu s$. The Transmon-based coupler is dynamically modulated on nanosecond timescales with coupling strength $ \cos{(\Delta_{ij} t+\varphi_{ij} )}$. The modulated phase $\varphi_{ij}$ is employed to synthesize the artificial magnetic field. The first sideband of the modulated frequency $\Delta_{ij}$ compensates for the energy gap between sites $i$ and $j$ during photon exchange. The photon hopping rate can reach up to 5 MHz.

	\textbf{State preparations.} The adiabatic path used to prepare the ground states has two steps: in the first step, the potentials of all sites keep unchanged and the coupling strengths gradually increase from 0 to 5 MHz; in the second step, the coupling strengths keep constant at 5 MHz and potentials of the disordered sites at the lattice center (boundary) gradually change from -15 MHz to 0 (from 0 to $ \pm V/2$). The change rates are optimized based on the system's energy gap $\Delta$. Within a time interval $dt$, the change $\delta$ is controlled as $\delta = c \cdot \Delta ^2 \cdot dt $. Here, $c \ll 1$ determines the total duration of adiabatic evolution 1 $\mu s$ (1.8 $\mu s$).
	
	\textbf{Data measurements.} 
	\textit{Butterfly spectrum}: To estimate the single-photon energy spectrum, we initialize one site $i$ in the lattice in a photon superposition state $ \left( {{{\left| 0 \right\rangle }_i} + {{\left| 1 \right\rangle }_i}} \right)/\sqrt 2 $ and turn on the couplings between the sites. We then measure the observables $ {\chi _i} = \left\langle {\hat \sigma _i^X} \right\rangle  + i\left\langle {\hat \sigma _i^Y} \right\rangle $ after different durations of evolution. The full energy spectrum is extracted by performing a Fourier transformation of ${\chi _i}$ for each index $i$ and average over all of them.
	
	\textit{Density correlation}: The photon density correlation function  is defined as the average spatial correlation between two photons as a function of their relative position ${\bf{d}}$, given by $ {g^{(2)}}({\bf{d}}) = \frac{N}{{N - 1}}\mathop \mu \limits_{{\bf{i,i + d}}} \frac{{\left\langle {{{\hat n}_{\bf{i}}}{{\hat n}_{{\bf{i + d}}}}} \right\rangle }}{{\left\langle {{{\hat n}_{\bf{i}}}} \right\rangle \left\langle {{{\hat n}_{{\bf{i + d}}}}} \right\rangle }} $, where  ${\hat n_{\bf{i}}}$ is the photon number on site ${\bf{i}}$ , and $N$ is the total photon number. The average, denoted as $\mu $, spans across all sites ${\bf{i}}$ with at least one site of ${\bf{i}}$ and ${\bf{i}} + {\bf{d}}$ belongs to the four central bulk sites of the system. We direct measure the photon density distributions to estimate the density correlation.
	
	\textit{Density current}: The photon density current is defined as the expectation value of two-site correlated operators, given by ${j_{(x,y),(x',y')}} = i{J}\left\langle {\sigma _{x,y}^{\dag }{\sigma _{x',y'}}} \right\rangle  + {\rm{H}}{\rm{.c}}{\rm{.}}$, where $(x,y)$ and $(x',y')$ represent the indices of two adjacent sites. We isolate the two measured sites by turning their connecting sites 1 GHz off resonance after preparing the ground state. We then fit the photon exchange pattern between the two sites with sinusoidal functions and extract the gradient at the beginning of isolation to estimate the density current.

	\textit{Hall conductivity}: The quantum Hall conductivity ${\sigma _{\rm{H}}}$ is defined as the variation of photon bulk density ${\rho _{{\rm{bulk}}}}$ with respect to the magnetic flux density using the Středa’s formula $\sigma _{\rm{H}} = \sigma _{\rm{0}} \cdot \frac{{\partial {\rho _{{\rm{bulk}}}}}}{{\partial (\varphi /2\pi )}} $, where ${\sigma _0} = 2e/h $. We measure the photon density distributions at different magnetic flux to extract the value of Hall conductivity.

	\textbf{Numerical simulations.} The numerical simulations in this work can be classified into three categories: exact diagonalization, pure state evolution under the system Hamiltonian $H$, and mixed state evolution under the Lindblad master equation with dephasing noise. The fidelities of the prepared states (gray solid line in Fig.~\ref{Fig3}d and Fig.~\ref{exfig1}), the changes of photon population (gray dashed line in Fig.~\ref{Fig5}b) and the response of photon bulk density (gray dashed line in Fig.~\ref{Fig5}c) represent the numerical simulation results under the Lindblad master equation, with an average dephasing time of $T_2=6.2 ~\mu s$ of each site. The photon deflection (solid line in Fig.~\ref{Fig2}a) and the butterfly spectrum (pseudocolor in Fig.~\ref{Fig2}b) represent the numerical simulation results of pure state evolution. The remaining numerical simulation results were obtained through exact diagonalization.
	
	~\\	
	~\\
	\textbf{Supplemental Figure}  	
	\renewcommand\thefigure{S\arabic{figure}}
	\setcounter{figure}{0}
	\begin{figure}[htbp] 		
		\includegraphics[width=0.9\linewidth]{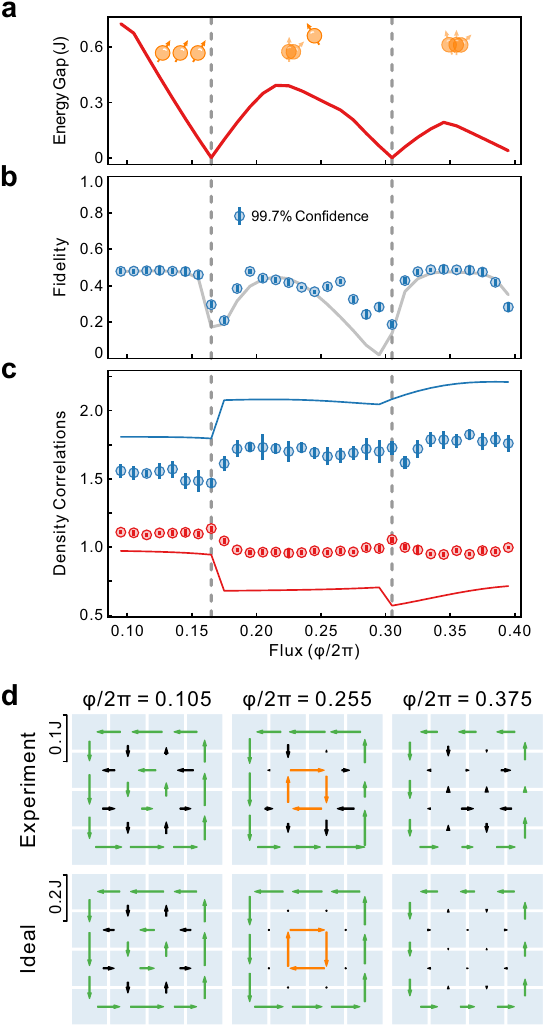} 		
		\centering 		
		\caption{\justifying{\textbf{Preparation and characterization of FQH states of N=3 photons.} \textbf{a,} Energy gaps as a function of magnetic flux. The gaps have two closing points, separating the phase into three parts: normal, crossover, and 3-photon FQH regime. In the new crossover regime, the lowest Landau level is likely to host two photons, similar to the 2-photon FQH state, while leaving the other photon in the higher level. \textbf{b,} Ground state fidelities are measured with respect to the magnetic flux. Two breakdowns in fidelity present at the phase transition points.  The gray solid line represents the simulation result with decoherence. \textbf{c,} The density correlations of small and large distances at varying magnetic flux. When transition from normal to FQH states, the density correlations are suppressed at small distances while being enhanced at larger distances. The solid lines represent the exact values obtained from diagonalization. \textbf{d,} The chiral density currents of the normal,crossover,and 3-photon FQH state, respectively. The crossover state shows a similar feature with the 2-photon FQH state (Fig.~\ref{Fig4}f). At 3-photon FQH state, the bulk currents and the radial currents flow between the boundary and the bulk are significantly suppressed. }}		
		\label{exfig1} 	
	\end{figure}

\end{document}